\documentclass[12pt]{article}
\usepackage{epsfig}

\usepackage{amssymb} 
\usepackage{amsmath} 
\usepackage{amsfonts} 
\usepackage{cite} 

\begin{document}
\newcommand{\wid}{0.9\columnwidth}
\newcommand{\beq}{\begin{equation}}
\newcommand{\eq}{\end{equation}}
\newcommand{\rb}{\underline{r}}
\newcommand{\kb}{\underline{k}}
\newcommand{\lb}{\underline{l}}
\newcommand{\qb}{\underline{q}}
\newcommand{\ab}{\underline{a}}
\newcommand{\bb}{\underline{b}}
\newcommand{\pb}{\underline{p}}
\newcommand{\Ab}{\underline{A}}
\newcommand{\Bb}{\underline{B}}
\newcommand{\Kb}{\underline{K}}
\newcommand{\Rb}{\underline{R}}
\newcommand{\zb}{\bar{z}}
\newcommand{\qqb}{\bar{q}}
\newcommand{\aab}{\bar{a}}
\newcommand{\lr}{\leftrightarrow}
\newcommand{\eps}{\epsilon}
\newcommand{\intfeyn}{\int\limits_0^1}
\def\be{\begin{equation}}
\def\ee{\end{equation}}
\def\bea{\begin{eqnarray}}
\def\eea{\end{eqnarray}}
\newcommand{\as}{\bar\alpha_s}
\begin{titlepage}

\begin{flushright}
\begin{tabular}{l}
 CPHT-PC 064.1105 \\
LPT 05-77 \\
 hep-ph 0511144n
\end{tabular}
\end{flushright}
\vspace{1.5cm}

\begin{center}

{\LARGE \bf Hard Pomeron in exclusive meson production at ILC
\footnote{Presented at Photon2005, International Conference on 
the Structure and Interactions of the Photon, Warsaw 31.08-04.09.2005, by Samuel Wallon.}%
}

\vspace{1cm}

{\sc R. ENBERG}${}^{a,b}$,
{\sc B.~Pire}${}^{a}$,
{\sc L.~Szymanowski}${}^{c,d}$ 
{\sc S.~Wallon}${}^{e}$
\\[0.5cm]
\vspace*{0.1cm} ${}^a${\it
CPhT, {\'E}cole Polytechnique, F-91128 Palaiseau, France\footnote{
  Unit{\'e} mixte 7644 du CNRS.} \\[0.2cm]
  \vspace*{0.1cm} ${}^b$
Lawrence Berkeley  Laboratory, Berkeley, Ca, USA\footnote{
  present address} \\[0.2cm]
\vspace*{0.1cm} ${}^c$  
 So{\l}tan Institute for Nuclear Studies,
Ho\.za 69, 00-681 Warsaw, Poland
                        \\[0.2cm]
\vspace*{0.1cm} ${}^d$ 
Universit\'e  de Li\`ege,  B4000  Li\`ege, Belgium  \\[0.2cm]
\vspace*{0.1cm} ${}^e$
LPT, Universit\'e Paris-Sud, F-91405 Orsay, France\footnote{
  Unit{\'e} mixte 8627 du CNRS.} \\[1.0cm]
}

\end{center}
\vskip2cm

We calculate  the exclusive process $\gamma^*_L  
(Q_1^2)\gamma^*_L(Q_2^2) \to
\rho^0_L \rho^0_L,$ at high energy.  The Born order estimate and the  
leading (LLA) and next to leading order (NLLA)
  BFKL resummation effects show the feasibility of
experimental detection  in  a quite large range of  
$Q^2$ values
at future high energy  $e^+e^-$ linear colliders.
\vskip1cm
\vspace*{1cm}

\end{titlepage}

\section{Introduction}
Future
 $e^+e^-$ colliders will offer the possibility of
clean testing of QCD dynamics through the scattering of two  virtual  
photons.
  By selecting events in which two vector mesons are produced
with a large rapidity gap and no accompanying particles,
one is getting access to the kinematical regime in which the  
perturbative
approach to exclusive scattering is justified. When
  the photon virtualities are comparable, the perturbative Regge  
dynamics of QCD
  should dominate and allow the use of resummation techniques  of the  
BFKL  type.
We  have studied \cite{PSW} these effects for the reaction
  \begin{equation}
\label{process}
\gamma^*_L(q_1)\;\gamma^*_L(q_2) \to \rho^0_L(k_1)  \;\rho^0_L(k_2)\,,
\end{equation}
where the virtualities  $Q_i^2 =-q_{i}^2$
  of the scattered photons play the role of the hard
scales.
Up to now, the available
 experimental data are restricted at rather small values of the  
energy
\cite{L3} and are analysed  in terms of generalized distribution  
amplitudes
\cite{gda}.
Recently the same process in the forward limit was studied in the full
NLLA BFKL approach in Ref.\cite{dima}.

  \section {Born order study}
The Born approximation study of the scattering amplitude of the process  
(\ref{process})
  proves the feasibility of a dedicated experiment.
  The impact representation of the scattering amplitude  has the form,
in terms of transverse momenta,
\begin{equation}
\label{M}
{\cal M}\! = \!is \!\int\!\frac{d^2\,\kb}{(2\pi)^4\kb^2\,(\rb -\kb)^2}
{\cal J}^{\gamma^*_L(q_1) \to \rho^0_L(k_1)}(\kb,\rb -\kb)
{\cal J}^{\gamma^*_L(q_2) \to \rho^0_L(k_2)}(-\kb,-\rb +\kb)\,,
\end{equation}
where ${\cal J}^{\gamma^*_L(q_i) \to \rho^0_L(k_i)}(\kb,\rb -\kb)$
are the impact factors corresponding  to the  transition of
$\gamma^*_L(q_i)\to \rho^0_L(k_i)$  via the
$t-$channel exchange of two gluons, with $\rb=1/2(\kb_1-\kb_2)$ and $\rb^2 = t_{min}-t\,.$
In this approximation, the  
amplitude (\ref{M}) depends linearly on $s$ and can be expressed as
\bea
\label{MCalgeneral}
  {\cal M} \!=i s 2 \pi \frac{N_c^2-1}{N_c^2} \alpha_s^2 \alpha_{em}
  f_\rho^2 Q_1 Q_2 
\intfeyn \! d z_1  d z_2 \,  z_1
 \zb_1 \, \phi(z_1) z_2
 \zb_2 \, \phi(z_2)\, {\rm M}(z_1,\, z_2)\,,
\eea
with (denoting $\mu_i^2=Q_i^2 z_i \, \zb_i$)
\bea
\label{defM}
&&{\rm M} (z_1,\, z_2) =  
\int \!\! \frac{d^2 \kb}{\kb^2 (\rb-\kb)^2} \!\!\left[ \frac{1}{z_1^2\rb^2 +  
\mu_1^2} +
\frac{1}{\zb_1^2\rb^2 + \mu_1^2}
  - \frac{1}{(z_1\rb -\kb)^2 + \mu_1^2}\right. \nonumber \\
&&\left.- \frac{1}{(\zb_1\rb
-\kb)^2
+ \mu_1^2} \right] \left[\frac{}{} (1 \leftrightarrow 2) \right]\,.
\eea
   We evaluated ${\rm M} (z_1,\, z_2)$ analytically and checked 
that the production amplitude dramatically decreases with  
$t$ so that its magnitude at
$t=t_{min}$  dictates the rate of the reaction. In this simpler case,  
the integral over $\kb$ can easily be performed and gives
\beq
\label{Mlimite}
M(z_1,z_2) = \frac{4 \pi}{z_1 \, \zb_1 z_2 \, \zb_2 Q_1^2 \, Q_2^2\,
(z_1 \, \zb_1 Q_1^2 - z_2 \, \zb_2 Q_2^2)} \ln \frac{z_1 \, \zb_1  
Q_1^2}{z_2 \, \zb_2 Q_2^2}\,.
\eq
The amplitude ${\cal M}$  can then be analytically computed
  through  $z_1,$ $z_2$ integration.
In the special case where $Q=Q_1=Q_2,$ it simplifies  to
\beq
\label{amplitudeR1}
{\cal M}_{t_{\min}}(Q_1=Q_2) = \, - \, i \,s \,  \frac{N_c^2-1}{N_c^2}  
\, \alpha_s^2 \, \alpha_{em} \,
  f_\rho^2 \, \frac{9 \pi^2}{2 Q^4}\, (24 - 28 \, \zeta(3))\,.
\eq
The peculiar limits $R=Q_1/Q_2 \gg 1$ and $R \ll 1$ correspond to the kinematics
typical for deep inelastic scattering on a photon target described through
collinear approximation. In the limit $R \gg 1$ the amplitude simplifies into
\beq
\label{amplitudeparton}
{\cal M}_{t_{\min}} \sim i s \frac{N_c^2-1}{N_c^2} \, \alpha_s^2 \, \alpha_{em} \,
\alpha(k_1) \, \beta(k_2) \, f_\rho^2 \, \frac{96 \pi^2}{Q_1^2\, Q_2^2}
\left(\frac{\ln R}{R}-\frac{1}{6 R} \right)\,.
\eq
We show on Fig.1L the differential cross section
${d \sigma/dt}=|{\cal M}|^2/ (16 \pi \,s^2 )$
at the threshold $t=t_{min}$ as a function of $Q^2_{2}/Q^2_{1}$.  
Its rather large 
order of magnitude 
when $Q_{1}^2 \approx Q_{2}^2 \approx 1\!-\!10 $ GeV$^2$ implies 
the feasibility of 
its experimental measure at the International Linear  
Collider.

\begin{figure}[htb]
\hspace{-.05cm}\epsfxsize=12.6cm{\centerline{\epsfbox{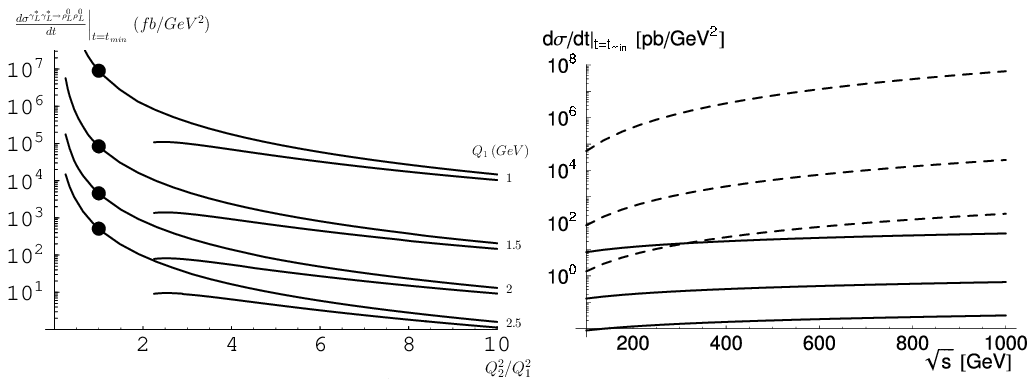}}}
\caption{{\it Left:} Differential cross-section
  for the process $\gamma^*_L\gamma^*_L \to \rho^0_L\rho^0_L$
at Born order, at the threshold $t=t_{min},$ as a function of  
$Q_2^2/Q_1^2.$
  The asymptotical curves, given by Eq.(\ref{amplitudeparton}), are valid for large
  $Q_2^2/Q_1^2$. The dots correspond to the special case $Q_1=Q_2$ where 
the formula (\ref{amplitudeR1}) can be applied.
{\it Right:} Cross section for LLA BFKL (dotted curves)  and for the NLLA
corrected kernel (full curves), for the three cases $Q=Q_1=Q_2=$ 2 GeV, 3 GeV and 4 GeV (from top to bottom in the plot).}   
\label{Figtmin}
\end{figure}

\section {BFKL effects}
We studied BFKL effects in the leading order approximation at  
$t=t_{min}\sim -Q_1^2 Q_2^2/s$.
In the BFKL framework, the amplitude of the process  can be expressed  
through the inverse Mellin transform with
respect to  $s$  as
\begin{equation}
A(s,t=t_{min}) = is \int\frac{d\omega}{2\pi i} e^{\omega Y}  
f_\omega(\rb^2=0)\;,
\label{ampldef}
\end{equation}
where   $Y=\ln (s/s_0)$ is the rapidity.
The BFKL Green's function reads
\bea
&&f_\omega(0) = \frac{1}{2 (2 \pi)^4}\int
\frac{dk^2}{k^3}\frac{dk'^2}{k'^3}
{\cal J}^{\gamma^*_L(q_1) \to \rho^0_L(k_1)}(\kb, -\kb)
{\cal J}^{\gamma^*_L(q_2) \to \rho^0_L(k_2)}(-\kb',\kb') \nonumber \\
&& \times \int_{-\infty}^{\infty} d\nu\;
\frac{1}{\omega - \omega(\nu)}\,
\left(\frac{k^2}{k'^2}\right)^{i\nu} \;,
\label{bfklamp3}
\eea
where the integration over angles has been performed.  
  $\omega(\nu)$
is the BFKL characteristic function  defined as
$\omega(\nu)=\as \chi(\nu)$,
with $\as \equiv \alpha_s N_c /\pi$ and
\beq
\label{defch}
\chi(\nu)= 2 \, \psi(1)-\psi\left(1/2 +i \nu
\right)-\psi\left(1/2 -i \nu \right)\,, \quad
\psi(x)=\Gamma'(x)/\Gamma(x).
\ee

At LLA accuracy, $\as$ is a fixed parameter.
  We choose to, however, let it depend on the given $Q_1$ and $Q_2$,  
which are external to the pomeron but provide
  a reasonable choice, through $\as =\frac{N_c}{\pi} \alpha_s(Q_1 Q_2)$.

To estimate higher order effects,  two improvements to the
LLA
BFKL amplitude have been made: BLM scale fixing (for the running of the coupling  entering the impact factors) and renormalization group resummed 
 BFKL kernel.

We show on Fig.1R the differential 
cross-section corresponding to the LLA BFKL and NLLA corrected
kernel.
 The amplification factor due to LLA BFKL  
resummation is indeed large. 
It is larger than the  enhancement  predicted for the total
cross-section \cite{bfklinc}, because the enhancement at the level of
amplitude
needs to be squared for exclusive processes. This effect is reduced 
 when including NLL corrections.

\section{Conclusion}
Since we expect the ILC  to cover a quite large  
region in
rapidity, experiments will allow us to test the dynamics of Pomeron  
exchange.
The Born approximation estimate  of this reaction, and the increase
of the amplitude due to BFKL resummation effects show that the process
$ \gamma^* \gamma^* \to \rho \rho $ should be measurable by
dedicated experiments at the ILC, for virtualities of  
the photons up to a few GeV$^2$.
We plan to pursue this line of research with Odderon exchange processes  
such as
  $ \gamma^* \gamma^* \to \pi^0 \pi ^0$ or, through
interference effects in charge asymmetric observables in
$ \gamma^* \gamma^* \to \pi^+ \pi^- \pi^+ \pi^- $ \cite{HPST}

\vskip.1in
\noindent
Work of L.Sz. is supported by the Polish Grant 1 P03B 028 28. 
He is a Visiting Fellow of the FNRS (Belgium).

\end{document}